# THE JANUS-FACE OF INNOVATION: GLOBAL DISPARITIES AND DIVERGENT OPTIONS

*This article examines how unequal access to AI innovation creates systemic challenges for developing countries. Differential access to AI innovation results from the acute competition between domestic and global actors. While developing nations contribute significantly to AI development through data annotation labor, they face limited access to advanced AI technologies and are increasingly caught between divergent regulatory approaches from democratic and authoritarian tendencies. This brief paper analyzes how more affordable AI engagement and Western countries' development cooperation present developing nations with a complex choice between accessibility and governance standards. I argue this challenge entails new institutional mechanisms for technology transfer and regulatory cooperation, while carefully balancing universal standards with local needs. In turn, good practices could help developing countries close the deepening gap of global technological divides, while ensuring responsible AI development in developing countries.*

## Nihat Muğurtay[*]

* Dr. Nihat Muğurtay is the PI of the TÜBİTAK-funded Global Diplomatic Network project and works as a post-doctoral researcher at the Computational Social Science Lab (VrlLab), Sabancı University.

Unequal access to AI innovation is not a future concern;[1] it is already contributing to developing countries' exposure to the technological divide and digital dependency. However, instead of reasoning about this puzzle, current debates on AI development reflect an alarmist attitude, ranging from national security concerns to domestic commercial competition among billion-dollar tech startups. This stems from a race among political and commercial actors to be the first in the AI market. However, such acute competition can lead to critical unintended spillovers for developing countries, which lag behind in AI innovation. With their growing populations and economies, developing countries will need AI-enhanced tools in many sectors for their social infrastructure and services. Unequal access to AI innovation is not merely a technological challenge, it is a systemic issue deepening global economic divides. In turn, this can accelerate digital dependency on more affordable (yet authoritarian) alternatives.

We should regard unequal access to AI innovation as a global externality generated by competition between companies and governments. Indeed, this externality can be managed through deeper collaboration between AI supplier governments, companies, and recipient countries. In this regard, whether technological development will create prosperity and greater equality depends on how we manage the institutional mechanisms that will shape its effects.[2] On the one hand, these effects will be shaped by developing countries' own agency and choices in how they participate in AI innovation and regulation. On the other hand, the international community, development partners, and private and public institutions will influence developing countries' active participation in AI innovation. Institutional mechanisms can foster specialized assistance programs, technology transfer agreements, collaborative research centers, and regional AI hubs through which governments and multilateral bodies can support the Global South in developing cost-effective AI solutions while maintaining appropriate regulatory standards. This approach would not only enhance their technological capabilities and close the tech divide, but also reduce their vulnerability to authoritarian misuse of AI innovation.

### *Annotating Inequality: Global South's Invisible Contribution Behind the AI Curtain*

People from developing countries have played an active, albeit ironic, role in AI development. However, their activity or agency reflects deep-rooted inequalities between high- and low-income countries. As a note, the power of many AI tools depends on the quality of annotated data or training processed by authentic human beings. Just as we eat fruits harvested by agricultural workers in developing countries, we utilize AI tools empowered by labor-intensive data annotation and labeling work from all over the world, including some of the world's least advantaged regions. This precarious data labor is being documented by witnesses, civil rights groups and unions.[3] A large number of people from the developing world are invited by top AI annotation platforms to improve AI tools. Indeed, it turns out that their pay range is much lower than that of any person from European countries.[4] This labor-intensive work of annotation makes AI tools more ethical and helps LLMs be more reliable in terms of instruction-following and truthfulness. Educated segments of developing countries label and classify images, traffic lights, pedestrians, and any visuals to inform AI tools.[5] This precarious part-time work has become a sector in Africa, and data annotation and AI trainer networks have already emerged.[6] However, African individuals' access to advanced AI tools are impeded by the brute fact of global inequalities. When a researcher, journalist or student from the developing world needs to use AI tools, tech companies' one-size-fits-all marketing strategy comes into play, namely these AI annotators and trainers pay the same price if they want to use advanced features of Unicorns' AI tools. This example depicts a paradox where the inclusion of people from the Global South simultaneously highlights their exclusion (or constraints). The presence of deep-rooted technological disparities pose a vital challenge for the Global South.

---

1) International Labour Organization. (2024). Mind the AI divide: Artificial intelligence and the global labor market. Retrieved from https://www.ilo.org/sites/default/files/2024-08/Mind%20the%20AI%20Divide_v12%20%281%29.pd
2) Acemoglu, D. & Johnson, S., & (2023). *Power and progress: Our thousand-year struggle over technology and prosperity*. Hachette UK.

Indeed, LLMs are just a minor glimpse of future debates. Differential access to advanced AI technologies will become increasingly significant as these tools are deployed in critical public and private sectors. Early projections by consultancy firms such as the PwC estimate that AI-driven technologies and investments have the potential to contribute up to $1.2 trillion to the African and Oceanian countries' economies, making 5.6 percent of their GDP.[7] However, UNDP underlines that Africa will receive only 10 percent of the global prosperity created by AI by 2030,[8] implying deeper inequalities and the digital divide bolstered by AI innovation. On this matter, not only private partnerships (imports or investments), but also foreign government funds and support -for deeper cooperation- will be critical. These options will create different sorts of exposure to AI facilities produced in foreign countries. I argue that

---

3) Okinyi, M. (2024). The hidden cost of training AI: Inside the lives of data labelers. Time
4) Mezzofiore, G., & Lomas, N. (2023). How OpenAI used Kenyan workers on less than $2 per hour to make ChatGPT better. Time. Retrieved from https://time.com/6247678/openai-chatgpt-kenya-workers/
5) Muldoon, J., Graham, M., & Cant, C. (2024, July 6). Meet Mercy and Anita – the African workers driving the AI revolution, for just over a dollar an hour. The Guardian.
6) Robertson, A., & Ghaffary, S. (2023, September 5). The hidden labor fueling AI's imagination. The Verge.
7) PwC. (2017). Sizing the prize: What's the real value of AI for your business and how can you capitalise? https://www.pwc.com/gx/en/issues/analytics/assets/pwc-ai-analysis-sizing-the-prize-report.pdf
8) United Nations Development Programme. (2024). Equitable AI for Africa.
Retrieved from https://www.undp.org/blog/equitable-ai-africa

developing countries' AI engagement with bilateral and multilateral development partners will play a pivotal role in shaping their ability to contribute to AI innovation, which, in turn, adheres to and implements standards, norms, and regulations related to AI innovation.

On the one hand, developing countries' engagement with capital-intensive countries can be facilitated by international trade. To close the tech divide, developing countries have their own alternative options and import AI-enhanced products that they need. AI products have a supply chain that includes production, deployment, and cross-country flows. Indeed, some countries such as China can offer affordable options for any AI tools in many sectors, and these are mostly directed to low and middle-income countries. Many questions arise like cross-border data flows, documentation and transparency, and friction between international regulations. China is conservative about not disclosing its financial flows (e.g. foreign aid), but its footprints are being tracked by various scholarly initiatives.[9] Companies like Bytedance (owner of TikTok), and its affiliates began annotating large amounts of visual and textual data as early as 2018.[10] In some other examples, scholars have documented China's export of surveillance technology to Serbia, a country currently seeking EU membership. Ghana is also one of the countries implementing the Safe City Project, which is powered by Huawei's artificial intelligence (AI) technology.[11] In total 64 countries imported the PRC's 155 AI projects between 2000 and 2017.[12] As the leading country in facial recognition technology, China has significant leverage to meet the affordable needs for public sectors.[13] In addition, China has been reluctant to join the export control regime as we saw in the Wassenaar Arrangement, a de-jure regime to prevent the dual use of high-tech exports. It is also essential to underline that surveillance technologies are not exclusively supplied by China; AI exporters include countries like France and other OECD members. Indeed, there are multiple layers of regulations limiting their exports. The EU institutions via its common commercial policy repeatedly raise concerns and impose regulations about the dual use of cyber-surveillance tools by importing governments.[14]

---

9) Bouey, J., Hu, L., Scholl, K., Marcellino, W., Dossani, R., Malik, A. A., Solomon, K., Zhang, S., & Shufer, A. (2023). China's AI Exports: Technology Distribution and Data Safety. RAND Corporation. https://www.rand.org/pubs/research_reports/RRA2696-2.html
10) Robertson, A., & Ghaffary, S. (2023, April 26). What is so special about TikTok's technology? Reuters. https://www.reuters.com/technology/what-is-so-special-about-tiktoks-technology-2024-04-26/
11) Roberts, T., Gitahi, J., Allam, P., Oboh, L., Oladapo, O., Appiah-Adjei, G., Galal, A., Kainja, J., Phiri, S., Abraham, K., Klovig Skelton, S., & Sheombar, A. (2024). Mapping the supply of surveillance technologies to Africa: Case studies from Nigeria, Ghana, Morocco, Malawi, and Zambia. Institute of Development Studies.
12) RAND Corporation. China's AI Exports Database (CAIED). https://www.rand.org/pubs/tools/TLA2696-1/tool.html
13) Madaio, M. (2022, April). How authoritarian regimes use artificial intelligence to suppress dissent. Harvard Magazine. https://www.harvardmagazine.com/2022/04/right-now-authoritarian-regimes-artificial-intelligence
14) Directorate-General for Trade. (2024, October 16). Commission publishes guidelines for cyber-surveillance exporters. European Commission.

The combination of loose export controls and lack of conditions on AI exports, along with weak institutional safeguards, makes developing countries more vulnerable to authoritarian (dual) uses of imported AI technologies. Unequal access to AI technologies and the growing global digital divide not only deepens AI dependency but also creates conditions that could strengthen unfettered misuse of AI. The absence of robust institutional governance mechanisms is becoming evident, as AI-driven online crimes and manipulations are on the rise. In Africa, Russian-linked disinformation campaigns are detected by social media corporations[15] and AI-generated fake audio clips sometimes spark political debates during electoral campaign periods.[16]

In addition to AI exports and imports, donor countries and multilateral organizations also provide AI-enhanced tools as foreign aid. Such grants can be budget support or any type of project, which has been used to promote recipients' development. However, foreign aid flows are not solely driven by recipient countries' needs - donors often aid based on their own political[17] and economic interests.[18] Although development aid provided by authoritarian countries can be positively associated with recipients' economic development,[19] the same logic does not necessarily apply in the political domain. When authoritarian donor-recipient pairs become deeply involved in AI through non-conditional foreign support, the expansion of surveillance—possibly reinforcing authoritarian practices—becomes more important. In other words, such divergent AI options are not only related to detecting tumors or helping students to learn foreign languages, but also related to how alternative AI options can transform societies into more progressive or degenerative paths or not. On this matter, developing countries' engagement with different cultures of AI innovation and regulation will be a significant element of their quest for responsible AI.

### *Divergent Cultures of AI Regulation*

The extent to which new technologies trigger inclusive or exclusionary regulatory tendencies depends on domestic institutional initiatives. On this matter, developing countries are not just passive bearers of AI innovation, but they also domestically and regionally find a way to make further steps to catch up to advanced countries. In July 2024, the African Union published its Continental AI Strategy and called member states to "develop national AI strategies … that emphasize building necessary capabilities to address the risks of AI and maximize its benefits." Indeed, African countries (Benin, Egypt, Ghana, Mauritius, Rwanda, Senegal, and Tunisia) build their own AI programs, yet not implementing regulations. The report also calls external development partners (e.g., foreign aid providers/private donors) to support African countries to establish their own AI infrastructure, which in turn, helps 'adoption for sustainable development goals (SDGs) and reduction of its risks to the society'. With this regard, developing countries' ability to comply with the human-centric AI ethics depend on their exposure to genuine linkages, reflecting different types and levels of AI engagement with global actors (e.g. governments and multilateral institutions). At this point, emerging cultures of AI become more apparent when it comes to AI regulation.

---

15) Mugurtay, N., Duygu, U., & Varol, O. (2024). Politics and Propaganda on Social Media: How Twitter and Meta Moderate State-Linked Information Operations. arXiv preprint arXiv:2401.02095.
16) Asiegbu, C., & Okolo, C. T. (2024, May 16). How AI is impacting policy processes and outcomes in Africa. Brookings.
17) Mugurtay, N., & Muftuler-Bac, M. (2022). Turkish Power Contestation with the United Arab Emirates: an Empirical Assessment of Official Development Assistance. *International Politics* (The Hague), 60(3), 659.
18) Muğurtay, N. (2022). Non-traditional donors and geopolitics of foreign aid: China, Turkey and the United Arab Emirates in Africa (Doctoral dissertation).
19) Mandon, P., & Woldemichael, M. T. (2023). Has Chinese aid benefited recipient countries? Evidence from a Meta-regression analysis. *World Development*, 166, 106211.

More democratic countries have bigger playing fields to trigger a public debate on the responsible use of AI. The EU, OECD and member states have long manifested an arena of contestation regarding human-centric AI innovation. In the United States, there has been an ongoing public debate on the use of AI-driven workplace surveillance tools.[20] This contest between employees and employers is accompanied by NGOs, journalists and researchers, meaning that there are diverse actors that can place themselves as active input providers for the responsible use of AI.[21] Despite democratic countries' varying levels of human-centric AI focus,[22] democratic countries possess the institutional safeguards to manage potential surveillance abuse. Indeed, citizens under authoritarian regimes can face much greater risks due to weaker protective mechanisms. This is the point that two emerging cultures of AI innovation and regulation come into play, and developing countries are caught between them.

G-7, the OECD and the EU offer multilateral arrangements for AI regulation. In February 2020, the OECD established the AI Policy Observatory (OECD. AI).[23] However, the EU is one of the actors that acted very early to regulate AI, indicating new technological tools' side-effects. The EU AI Act categorizes AI systems according to their risk levels, from minimal to unacceptable risk. On the other hand, China also is one of the early birds to regulate AI innovation. In this regard, the EU AI Act (2024) and Chinese AI regulation[24] are fruitful use cases to show how emerging cultures of AI regulation share differentiated objectives. For the former, AI regulations pay more attention to trustworthiness, criminal offenses, safety or human-security, suggesting different risk levels. Safety processes entail strong monitoring and auditing for human-security. On the other hand, Chinese precautions are also related to the promotion of socialism, greater political control, and follow the Chinese Communist Party's agenda.[25] Cyberspace Administration of China (CAC)—the responsible state apparatus for AI regulation—reiterates its socialist vision in the Interim Measures for the Management of Generative Artificial Intelligence Services (2023).[26]

---

20) The Week. (2024). Workplace AI surveillance, from https://theweek.com/tech/workplace-ai-surveillance
21) Hickok, M., & Maslej, N. (2023). A policy primer and roadmap on AI worker surveillance and productivity scoring tools. *AI and Ethics*, 3(3), 673-687.
22) Center for AI and Digital Policy. (2023). Artificial Intelligence and Democratic Values 2023. Retrieved from https://www.caidp.org/reports/aidv-2023/
23) See https://oecd.ai/en/
24) Other multilateral initiatives for AI guidelines, regulations and mandates are also offered by G7, G20, the OECD AI Principles (2019), UNESCO.

Generative AI is, again, a useful example to visualize these emerging cultures of AI regulation. Most of the large language models are already reflecting their creators' attitude in social and political affairs.[27] While some LLM models demonstrate a strong emphasis on liberal democratic principles including diversity, sustainability, and human rights, some others tend to emphasize state authority, traditional morality, and centralized governance. For instance, AI tools -trained and deployed- in China must meet certain ideological requirements related to more governmental control. This is currently visible in the case of the DeepSeek, a powerful LLM tool produced in China. When asked about topics such as democracy or historical events like the Tiananmen Square, it reflects China's domestic AI regulations by either refusing to answer or providing responses aligned with official narratives.[28] Moreover, DeepSeek is significantly more affordable compared to its Western counterparts, making it an attractive option for developing countries seeking advanced AI solutions. Chinese AI models like Qwen, produced by Alibaba, also have the potential to challenge the global LLM market with more affordable options.[29] Indeed, LLMs are just the tip of the emerging story. Developing countries find themselves caught between competing approaches to AI innovation emerging from different global powers. More enhanced and sophisticated AI tools will trigger acute debate to alleviate AI's side-effects. Indeed, this is not inevitable since fruitful policy frameworks are possible to aid the Global South in gaining capabilities in AI innovation.

### *Closing the Gap*

More steps have been taken to close the tech divide. Affordable AI trade and development cooperation can close the tech divide between developing countries and capital-intensive countries. Governmental programs -combined with the civil sector-can mobilize significant resources for the responsible use of AI. Capital-intensive countries and their private sector actors can collaborate to deal with closing the gap between industrialized and developing countries. The U.S. government, Amazon, Anthropic, Google, Meta, Microsoft and OpenAI came together and initiated a program to decrease global tech disparities.[30] However, these initiatives are still embryonic and providing free API keys via researcher access initiatives or other small grants does not satisfy what developing countries need. Indeed, actual progress on the ground and use of AI in development sectors will signal a robust common ground between development partners and recipient countries. Public-private partnerships in development cooperation will be a key issue for the benefit of developing countries.

---

25) China's AI Regulation (2022): https://digichina.stanford.edu/work/translation-internet-information-service-algorithmic-recommendation-management-provisions-effective-march-1-2022/
26) China Law Translate. Comparison chart of current vs draft rules for generative AI. Retrieved December 4, 2024, from https://www.chinalawtranslate.com/en/comparison-chart-of-current-vs-draft-rules-for-generative-ai/
27) Buyl, M., Rogiers, A., Noels, S., Dominguez-Catena, I., Heiter, E., Romero, R., ... & De Bie, T. (2024). Large Language Models Reflect the Ideology of their Creators. arXiv preprint arXiv:2410.18417.
28) McCarthy, S. (2025, January 29). DeepSeek is giving the world a window into Chinese censorship and information control. CNN. https://edition.cnn.com/2025/01/29/china/deepseek-ai-china-censorship-moderation-intl-hnk/index.html
29) Tech in Asia. (2024). Chinese AI models like Alibaba's Qwen rival OpenAI, Meta. Retrieved January 16, 2025, from https://www.techinasia.com/news/chinese-ai-models-alibabas-qwen-rival-openai-meta

We can trace the predecessors of such effective policies by examining actual developments on the ground. To illustrate, AI is being used and tested in multiple humanitarian and development sectors including disaster prevention.[31] Such AI-driven early warning systems or disaster coordination tools can decrease casualties and material damage. Another example shows how intelligent drones can detect early signs of agricultural diseases. Furthermore, a chatbot can help villagers to deal with their concrete problems in agricultural sectors in their local language.[32] In Nigeria, AI-driven healthcare facilities are being used by citizens.[33] For instance, Nigerian startup Ubenwa analyzes baby cries to diagnose birth asphyxia, a major cause of infant mortality in Africa.[34] Doing so, AI-enhanced tools can also touch a long-standing problem of aid effectiveness. Given the AI tools' promise for productivity, marginal returns -for social sectors- can be much higher than expected.[35] Such examples indicate the actual link between AI and development sectors, and help developing countries close the deep tech divide. These all have been made possible with aid coming from international private and public partners, including the World Bank and many others. In the future, we will see many sophisticated tools that support local communities' social and economic well-being.

These new developments (and possibly AI transformation) come with some downsides and limitations. Such good use-cases should be sustainable, not volatile. A useful example is Artificial Intelligence for Development (AI4D), a collaboration between local and global stakeholders. This partnership goes beyond providing technology - they help create governance frameworks that reflect African countries' diverse needs and conditions, by involving local stakeholders in policy planning. On this manner, effectiveness of AI to promote prosperity or progressive agenda depends on how domestic and global development partners manage AI innovation.[36] In other words, AI won't perform any magic or act as a Midas touch to solve the persistent puzzles in developing countries. In case of a lack of necessary governance mechanisms and regulatory vision, development partners (such as governments) can also take some precautions to prevent the dual use/misuse of sophisticated AI tools.

---

30) U.S. Department of State. (2024). Secretary Antony J. Blinken at the advancing sustainable development through safe, secure, and trustworthy AI event.
31) Kuglitsch, M. M., Cox, J., Luterbacher, J., Jamoussi, B., Xoplaki, E., Thummarukudy, M., ... & Ward, T. (2024). AI to the rescue: how to enhance disaster early warnings with tech tools. *Nature*, 634(8032), 27-29.
32) Chow, A. R. (2024, October 31). Inside the new nonprofit AI initiatives seeking to aid teachers and farmers in rural Africa. Time.
33) Bill and Melinda Gates Foundation (2023). Nneka Mobisson and mDoc: Transforming healthcare in Africa. https://www.gatesfoundation.org/ideas/articles/nneka-mobisson-mdoc-healthcare
34) See https://www.ubenwa.ai/news
35) OECD. (2024). The impact of artificial intelligence on productivity, distribution, and growth. Organisation for Economic Co-operation and Development.

### *Challenges of AI Regulation*

The implementation of AI regulations in developing countries faces unique challenges that stem from both technical and socio-political complexities. This becomes particularly evident when examining how traditional development cooperation handles governance challenges. The European Union and member states' development policy is a fruitful example of how to deal with governmental abuses and the misuse of foreign support. In the traditional sense, for instance, if recipient governments violate fundamental rights—whether through political crackdowns,[37] then the donors suspend their development cooperation with these recipient governments. In the case of AI, the use of advanced algorithms in public sectors also carries the risk of discrimination against certain groups or the oppression of minorities, often in ways that are more subtle.[38] Given that some of these algorithmic biases have already occurred in technologically advanced countries (which have also taken steps in AI regulations), the use of AI tools in environments with poor governance records demands greater attention. In a similar vein, when development partners provide AI-enhanced foreign support or export these tools for use in the education sector, it is essential to incorporate safeguards to prevent human rights violations. Furthermore, governments may use AI to promote citizens' well-being in the health sector, but development partners may also insist that AI-driven strategies adhere to ethical standards, such as preventing the misuse of AI tools on specific segments of society.

This regulation issue can also initiate another round of debate between divergent options from Global South. Moreover, aid conditionalities and selectivity frequently imposed by Western donors are often met with criticism from recipient countries. For instance, after Uganda's restrictive LGBT laws lead to foreign suspensions from the European Union, African leaders often addressed their local culture and faith while evaluating gender-related aid suspensions.[39] When it comes to Western-generated AI regulations, some similar voices from African communities send early warnings, addressing problems of applying Western AI regulations to African contexts.[40] Moreover, there are also voices addressing biases in the current AI tools, suggesting "Western datasets reinforce existing power structures and perpetuates a singular, dominant worldview."[41] Relatedly, other development partners also join the debate after suspensions. To illustrate, after Uganda's foreign aid was suspended by the EU, Russia-backed legal support teams came into play to support local government to act against such Western conditionalities.[42] As many developing countries have constructed ports and railways with the support of more affordable yet authoritarian partners, they can do the same to deal with the deep tech divide. Therefore, strict conditional support for AI development would risk creating tension between universal standards and local needs. Externally advised (or imposed) regulations (via bilateral donors or multilateral institutions) might unintentionally create alienation and push developing nations toward more accessible but potentially authoritarian AI alternatives. Local civil society groups, including NGOs and academic institutions, can play a critical role in addressing local views and potential harms and advocating for inclusive AI strategies that respond to on-the-ground realities. These are the minefields that development communities, stakeholders and policymakers will deal with in the near future.

---

36) Acemoglu, D. & Johnson, S., & (2023). Power and progress: Our thousand-year struggle over technology and prosperity. Hachette UK.

37) Reuters. (2023, July 29). EU suspends budget support and security cooperation with Niger. Reuters.

38) Bedayn, J. (2024). Class action lawsuit on AI-related discrimination reaches final settlement. AP News.

However, we must note that In AI exporter countries, these regulations are still a contested issue, and not yet complete or fully aligned with human rights and inclusivity principles. Recent indexes indicate industrialized countries are far from prioritizing harmonization between AI and human-centric innovation. AI regulations vary widely even among industrialized countries and cannot be reduced to a Manichean duality. Unicorns dominate the industry in the West, playing with large amounts of data, and using them in their own AI-driven tools. At the domestic level, tech startups drain the national battery to satisfy the massive energy needs for training AI models.[43] Commercial competition -as seen in the case of generative AI- has reached unprecedented levels. Moreover, each new technology brings new ideas on regulations, triggering a cycle of legal and regulatory efforts aiming to keep pace with innovation. Indeed, the game of catch-up between policymakers and AI innovation will continue. Therefore, the AI regulation debate will continue in the near future, and a balanced approach is necessary to address the delicacy of the tech divide and the potential misuse of AI innovation. During these processes, more voices from the developing world need to be taken into consideration, which, in turn, can create a more participatory and global AI environment.

---

39) Reuters. (2023, August 10). Uganda president defiant after World Bank funding suspended over LGBT law. Reuters.

40) Hook, F. (2024, February). The ethical considerations of AI in Africa. Connecting Africa. Retrieved January 8, 2025, from https://www.connectingafrica.com/ai/the-ethical-considerations-of-ai-in-africa

41) Tibebu, H. (2024, May 1). Why Africa must demand a fair share in AI development and governance. Tech Policy Press. Retrieved January 8, 2025, from https://www.techpolicy.press/why-africa-must-demand-a-fair-share-in-ai-development-and-governance/

42) Marinelli, C. (2024, September). A U.S. activist allegedly accepted $300K from Russia to push anti-gay laws in Africa. LGBTQ Nation.

43) The Economist. (2024, November 18). Will the bubble burst for AI in 2025, or will it start to deliver? https://www.economist.com/the-world-ahead/2024/11/18/will-the-bubble-burst-for-ai-in-2025-or-will-it-start-to-deliver

We also should note that uncertainties about tech divide and regulations can be mitigated by bilateral or multilateral diplomacy among countries. Different emerging approaches to AI innovation and regulation do not have to create barriers - instead, they can serve as grounds for meaningful dialogue between different systems.[44] The U.S.-China high-level diplomacy in AI innovation took place in May 2024.[45] Moreover, France and China released a joint declaration on AI, addressing risks of AI.[46] Technologically advanced countries -representing different cultures of AI regulation- can mitigate side-effects of acute competition, which in turn, can positively affect developing countries in the Global South. A global program aimed at enhancing the tech integration of developing countries into global politics could provide a unifying platform to address and mitigate future uncertainties.

---

44) West, D. M., & Allen, J. R. (2023, February 7). Laying the groundwork for US-China AI dialogue. Brookings Institution.
45) Associated Press. (2024, May 17). Top US and Chinese officials begin talks on AI in Geneva. AP News. https://apnews.com/article/artificial-intelligence-china-united-states-geneva-switzerland-1aa4451f82f250a47039a213f3d72879
46) Global Times. (2024, May 7). China, France release joint declaration on AI governance, agreeing to work closer. Global Times.